\documentclass[paper]{JHEP3} % 10pt is ignored!

\usepackage{epsfig,multicol}
\usepackage{amsmath}
\usepackage{amscd}
\usepackage{amsfonts}
\usepackage{amsthm}

\def\IN{\mathbb{N}}

\def\IR{\mathbb{R}}
\def\ID{\mathbb{D}}

\def\ap#1{\alpha^{\prime\,#1}}

\newcommand\fverb{\setbox\pippobox=\hbox\bgroup\verb}
\newcommand\fverbdo{\egroup\medskip\noindent%
            \fbox{\unhbox\pippobox}\ }
\newcommand\fverbit{\egroup\item[\fbox{\unhbox\pippobox}]}
\newbox\pippobox
\newcommand{\Li}{\operatorname{Li}}

\title{Derivative corrections to the Born-Infeld action through beta-function calculations in
N=2 ``boundary'' superspace}

\author{Stijn Nevens, Alexander Sevrin\\
Theoretische Natuurkunde, Vrije Universiteit Brussel and\\The International Solvay Institutes\\
Pleinlaan 2, B-1050 Brussels, Belgium \\
E-mail:  \email{stijn@tena4.vub.ac.be}, \email{Alexandre.Sevrin@vub.ac.be}}
\author{Walter Troost\\
Instituut voor Theoretische Fysica,
Katholieke Universiteit Leuven,\\
Celestijnenlaan 200D
B-3001 Leuven, Belgium.\\
E-mail:  \email{Walter.Troost@fys.kuleuven.ac.be}}
\author{Alexander
Wijns\thanks{Aspirant FWO}\\
Theoretische Natuurkunde, Vrije Universiteit Brussel and\\The International Solvay Institutes\\
Pleinlaan 2, B-1050 Brussels, Belgium \\
E-mail:  \email{awijns@tena4.vub.ac.be}}

\preprint{\hepth{0606255}}  % OR: \preprint{Aaaa/Mm/Yy\\Aaa-aa/Nnnnnn}
\abstract{We calculate the $ \beta $-functions for an open string $ \sigma $-model in the
presence
of a $U(1)$ background. Passing to $N=2$ boundary superspace, in which the background is fully
characterized by a scalar potential, significantly facilitates the calculation. Performing the
calculation through three
loops yields the equations of motion up to five derivatives on the fieldstrengths, which upon
integration gives the bosonic sector of the effective action for a single D-brane in trivial
bulk background fields through four derivatives and to all orders in $ \alpha {}'$. Finally, the
present calculation shows that demanding ultra-violet finiteness of the non-linear $ \sigma
$-model can be reformulated as the requirement that the background is a deformed
stable holomorphic $U(1)$ bundle.}


\begin{document}
\setcounter{equation}{0}

%
% Introduction
%

\section{Introduction} \label{introduction}
The effective world volume action for $n$ coinciding Dp-branes is, in leading order in
$ \alpha '$, given by the
$d=9+1$, $N=1$ supersymmetric $U(n)$ Yang-Mills action dimensionally reduced to $p+1$
dimensions \cite{witten}. For a single D-brane, $n=1$, the effective action is known to all
orders in $ \alpha '$ in the limit of constant (or slowly varying) background fields: it is the
$d=9+1$, $N=1$ supersymmetric Born-Infeld action, dimensionally reduced to $p+1$
dimensions, \cite{BI1} -- \cite{BI7}. Both bosonic and fermionic terms as well as the couplings
to the
bulk background fields are known. Derivative corrections were studied in
\cite{andreevtseytlin} (using the partition function method), in \cite{wyllard} (using boundary
conformal field theory) and \cite{cornalba} (using the Seiberg-Witten map). Modulo field
redefinitions, it was shown that there are no two derivative corrections and a proposal for the
four derivative
corrections through all orders in $ \alpha '$ was made \cite{wyllard}.
Supersymmetric extensions of the derivative corrections have been studied as well
\cite{Collinucci:2002gd} (for some other supersymmetry inspired considerations see e.g. %%@
\cite{goteborg}).

For $n>1$, the situation is more involved. Using the symmetrized trace description, a
non-abelian generalization of the Born-Infeld has been proposed \cite{Tseytlin:1997cs}. However,
this can not be the full answer (as was in fact stressed in \cite{Tseytlin:1997cs}), as the
non-abelian Born-Infeld action defined in this way does not correctly reproduce the mass
spectrum of strings stretched between intersecting branes \cite{Hashimoto:1997gm},
\cite{Denef:2000rj}. In fact, this action does not even
allow for a supersymmetric extension \cite{bdrs}.
Ignoring derivative corrections is equivalent to requiring
that the background fields are constant. This leads, because of
$D_aF_{bc}=0\Rightarrow [F_{ab},F_{cd}]=0$, back to the abelian situation. So here, one has
to deal with derivative corrections right from the start. Only partial results are known.
Indeed, there are no
$ {\cal O}( \alpha ')$ corrections and the $ {\cal O}( \alpha '{}^2)$ corrections were
calculated from
open superstring amplitudes in \cite{direct}. Calculating higher order contributions from string
scattering amplitudes is very involved\footnote{However note that the $ \alpha {}'{}^3$ results
of \cite{alpha3} were verified by a string calculation, \cite{Barreiro:2005hv}.
The results of \cite{stieberger} show
that at higher order even the impossible might be possible!}. As a consequence,
alternative methods to construct the effective action have been developed. Requiring that %%@
certain BPS
configurations solve the
equations of motion \cite{lies} allowed one to calculate both the $ \alpha '{}^3$ \cite{alpha3},
and the
$ \alpha '{}^4$ \cite{alpha4}, corrections (see also the summarizing equations in
\cite{testalpha4}). An alternative method, albeit confined to four dimensions, was
developed in \cite{Grasso:2002wb}, \cite{Refolli:2001df} and agreed with the results
in \cite{alpha3}.
While very explicit, these results are very involved and not particularly
illuminating.

This strongly suggests that one might first want to control the derivative corrections in the
-- much simpler --
abelian case, possibly obtaining all order expressions. A direct calculation through string
scattering amplitudes is very hard. Past experience with the calculation
of the $ \alpha '$-corrections
to the bulk equations of motion showed that the calculation of $ \beta $-functions in the
corresponding
non-linear $ \sigma $-model is a particularly powerful approach, \cite{Grisaru1}.
When reformulated in $N=(2,2)$ superspace, the calculation greatly simplifies, \cite{Grisaru2}.

In the present paper we calculate the $ \beta $-functions for an open string $ \sigma $-model in
the presence of a
$U(1)$ background field\footnote{Note that $ \beta $-function calculations (through two loops)
were used to gain insight in the bulk/brane couplings \cite{Bordalo:2004xg}.}. We perform the %%@
calculation in the $N=2$ boundary superspace developed
in \cite{Koerber:2003ef}
where many of the simplifying features discovered in \cite{Grisaru2} persist.

The paper is organized as follows. In the next section we set up the $ \sigma $-model in $N=2$
boundary superspace. This is followed by an analysis of the calculation to be performed and a
derivation of the necessary ingredients such as the superspace propagators. In section 4 we
focus on the $ \beta $-functions at one and two loops. These calculations are reproduced in
detail in appendix B. Section 5 turns to the three loop calculation. Again we refer the reader
interested in technical details to appendix C. We end with our conclusions. Conventions and a
useful diagrammatic representation are developed in appendix A.

\section{$N=2$ non-linear sigma model with boundaries in $N=2$ ``boundary'' superspace}
\label{N2NonLinSigwBinBS}
In this section we present the action for an open string $ \sigma $-model in a $U(1)$
background.
The treatment is greatly simplified if we formulate the model in $N=2$ boundary
superspace. Indeed, the whole $U(1)$ structure turns out to be characterized by a single
scalar potential $V$. The model at hand is a special case of the general setup developed
in \cite{Koerber:2003ef}.

We introduce chiral fields, $Z^ \alpha $, and anti-chiral fields ,
$Z^{\bar \alpha }$, $ \alpha \in\{1,\cdots m\}$, satisfying the
constraints\footnote{Our conventions are given in appendix \ref{app
conv}.},
\begin{eqnarray}
\bar D Z^ \alpha =D Z^{\bar \alpha }=0.\label{con1}
\end{eqnarray}
In addition we need a set of fermionic constrained fields $ \Psi^ \alpha  $ and $ \Psi ^{\bar
\alpha }$, which satisfy,
\begin{eqnarray}
\bar D \Psi ^ \alpha = \partial _ \sigma  Z^ \alpha ,\qquad
D \Psi ^{\bar \alpha }= \partial _ \sigma Z^{\bar \alpha }.\label{con2}
\end{eqnarray}
The action, $ {\cal S}$, consists of a free bulk term, $ {\cal S}_0$,
and a boundary interaction term, $ {\cal S}_{int}$,
\begin{eqnarray}
{\cal S}= {\cal S}_0+ {\cal S}_{int},\label{nlsm1}
\end{eqnarray}
where,
\begin{eqnarray}
{\cal S}_0&=&  \int d^2 \sigma d^2 \theta \left(
g_{ \alpha \bar \beta }\,DZ^ \alpha \bar DZ^{\bar \beta }+
g_{ \alpha \bar \beta }\, \Psi ^ \alpha \Psi ^{\bar \beta }
\right), \nonumber\\
{\cal S}_{int}&=&-\int d \tau d^2 \theta \, V\left( Z, \bar Z\right),\label{nlsm2}
\end{eqnarray}
where the potential $V$ in $ {\cal S}_{int}$ is at this point an arbitrary real function
of the chiral and the anti-chiral superfields.
In these equations, we rescaled $Z$ (and $ \Psi $) by a factor
$\sqrt{2 \pi  \alpha '}$ such as to make
it dimensionless (of dimension $1/2$ resp.). We choose Neumann boundary conditions,
\begin{eqnarray}
\Psi^ \alpha  \Big|{}_{\mbox{boundary}}=\Psi^{\bar \alpha } \Big|{}_{\mbox{boundary}}=0.
\label{bdycds}
\end{eqnarray}
The boundary term gives the coupling to a $U(1)$ background field. The magnetic fields, %%@
appropriately
rescaled by a factor $2 \pi  \alpha '$, are obtained from the potential,
\begin{eqnarray}
&&F_{ \alpha \bar \beta }= i V_{ \alpha \bar \beta } \nonumber\\
&&F_{ \alpha \beta }=F_{\bar \alpha \bar \beta }=0,\label{fspot}
\end{eqnarray}
where here and in the next we use the notation,
\begin{eqnarray}
V_{ \alpha _1\cdots \alpha _m \bar \beta _1\cdots \bar \beta _n} \equiv
\partial _{\alpha _1}\cdots \partial _{\alpha _m}
\partial _{ \bar \beta _1}\cdots \partial _{\bar \beta _n} V.
\end{eqnarray}
Whenever needed we take the potentials to be,
\begin{eqnarray}
A_ \alpha =-\frac i 2 \partial _ \alpha V, \qquad A_{ \bar \alpha }
=\frac i 2 \partial _{ \bar \alpha }V.\label{pots}
\end{eqnarray}

We will treat the potential using a background field expansion: we take the superfields to be %%@
the sum
of some solution of the equations of motion with the quantum fluctuations.
Expanding the potential around the classical configuration, we get the
interactions,
\begin{eqnarray}
&&{\cal S}_{int}=-\int d \tau d^2 \theta \left(
\frac 1 2 V_{ \alpha \beta }Z^ \alpha Z^ \beta + \frac 1 2 V_{ \bar \alpha \bar \beta }
Z^{ \bar \alpha }Z^{ \bar \beta }+ V_{ \alpha \bar \beta }Z^ \alpha Z^{ \bar \beta }+
\right.\nonumber\\
&&\quad
\left. \frac{1}{3!}V_{ \alpha \beta \gamma }Z^{ \alpha }Z^{ \beta }Z^{ \gamma }+
\frac{1}{3!}V_{ \bar  \alpha \bar \beta \bar \gamma }Z^{ \bar  \alpha }Z^{ \bar \beta }
Z^{ \bar \gamma }+
\frac 1 2 V_{ \alpha \beta \bar \gamma }Z^{ \alpha }Z^{ \beta } Z^{ \bar \gamma }
+\frac 1 2 V_{ \bar  \alpha \bar \beta \gamma }Z^{ \bar  \alpha }Z^{ \bar  \beta } Z^{\gamma }
+\cdots
\right), \label{bgint}\nonumber\\
\end{eqnarray}
where the vertices are functions of the background fields and the quantum fluctuations are
denoted by $Z$.
The terms linear in the fluctuations vanish because the background fields solve the equations
of motion.

Before turning to the actual calculations, we will first analyze what to expect. The bare %%@
potential
$V_{bare}$ will be of the form,
\begin{eqnarray}
V_{bare}=V+\sum_{r\geq 1} V_{(r)} \lambda ^r,
\end{eqnarray}
with $ \lambda = (\ln \, \Lambda) /\pi$ and $ \Lambda =M/m $ with
$M$ the UV cut-off and $m$ the IR regulator. Making the loop
expansion explicit we get
\begin{eqnarray}
V_{(r)}=\sum_{s\geq r} V_{(r,s)}\,\hbar^s. \label{loopexp}
\end{eqnarray}
Using eq.~(\ref{pots}), one finds that the $ \beta $-functions for
the gauge potentials are then given by,
\begin{eqnarray}
\beta _ \alpha =\frac{i}{2\pi} \partial _ \alpha V_{(1)}, \qquad
\beta _ { \bar \alpha} =-\frac{i}{2\pi} \partial _ { \bar \alpha}
V_{(1)}.\label{bfundef}
\end{eqnarray}
The renormalization group recursively expresses $V_{(r)}$, $r\geq 2$, in terms of $V_{(1)}$,
\begin{eqnarray}
V_{(r+1)}(A)=- \frac{\pi}{r+1}\, V_{(r)}(A+ \beta
)\Big|_{\mbox{\small part linear in } \beta } . \label{rg}
\end{eqnarray}
This fact provides a strong consistency check on the loop calculations.

\section{Propagators, vertices and superspace technology}
In order to calculate the propagator we introduce unconstrained sources $J$, $\bar J$, $ \Omega$ %%@
and
$\bar \Omega$. Adding them to the free action,
\begin{eqnarray}
{\cal S}=  \int d^2 \sigma d^2 \theta \left(
g_{ \alpha \bar \beta }DZ^ \alpha DZ^{\bar \beta }+
g_{ \alpha \bar \beta } \Psi ^ \alpha \Psi ^{\bar \beta }+
J_ \alpha Z^ \alpha +\bar J_{\bar \alpha }Z^{\bar \alpha }+ \Omega_ \alpha \Psi ^ \alpha
+\bar \Omega _{ \bar \alpha } \Psi ^{\bar \alpha }
\right),
\end{eqnarray}
we get, upon completing the squares, the propagators,
\begin{eqnarray}
\langle Z^ \alpha (1) Z^{\bar \beta }(2)\rangle&=&- g^{ \alpha \bar \beta }
\frac{D_2\bar D_2}{\Box_2 } \delta ^{(4)}(1-2), \nonumber\\
\langle \Psi ^ \alpha (1)\Psi ^{\bar \beta }(2) \rangle &=&
-g^{ \alpha \bar \beta } \left( 1-\bar D_2D_2 \frac{ \partial _{ \tau _2}}{\Box_2}
\right)\delta ^{(4)}(1-2),\nonumber\\
\langle Z ^ \alpha (1)\Psi ^{\bar \beta }(2) \rangle &=& -g^{ \alpha \bar \beta }
\bar D_2 \frac{ \partial _{ \sigma  _2}}{\Box_2}
\delta ^{(4)}(1-2),\nonumber\\
\langle \Psi ^ \alpha (1) Z ^{\bar \beta }(2) \rangle &=&g^{ \alpha \bar \beta }
D_2 \frac{ \partial _{ \sigma _2}}{\Box_2}
\delta ^{(4)}(1-2) ,\label{fullprops}
\end{eqnarray}
which satisfy the boundary conditions eq.~(\ref{bdycds}). As a consistency check, one verifies %%@
that
the propagators are compatible with the constraints, eqs.~(\ref{con1}) and (\ref{con2}).

For the calculation at hand, we only need the first of the propagators in eq.~(\ref{fullprops})
evaluated at the boundary. It becomes,
\begin{eqnarray}
\ID{}^{ \alpha \bar \beta }(1-2)&\equiv&
\langle Z^ \alpha (1) Z^{\bar \beta }(2)\rangle\Big|{}_{\mbox{\small boundary}} \nonumber\\
&=&
g^{ \alpha \bar \beta }
D_2\bar D_2 \left(\frac 1 \pi \int_m^M \frac{d\,p}{p}\,\cos\big(p\,( \tau _1- \tau _2)\big)\, %%@
\delta ^{(2)}( \theta _1- \theta _2)\right) \nonumber\\
&=&-g^{ \alpha \bar \beta }
\bar D_1 D_1 \left(\frac 1 \pi \int_m^M \frac{d\,p}{p}\,\cos\big(p\,( \tau _1- \tau _2)\big)\, %%@
\delta ^{(2)}( \theta _1- \theta _2)\right),\label{propone}
\end{eqnarray}
where the IR regulator was denoted by $m$ and the UV cut-off by $M$.

\begin{figure}[h]
\begin{center}
\psfig{figure=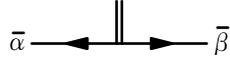}%,height=0.3in}
\caption{A diagram containing vertices with only holomorphic or only anti-holomorphic indices %%@
will
be UV finite.
\label{fig:vertex}}
\end{center}
\end{figure}

When calculating the $ \beta $-functions, we are solely interested in the UV divergences which
we will
treat using minimal subtraction. The actual calculation is further simplified by several
observations made in \cite{Grisaru2} which carry over to
the present case. One verifies that all UV divergences are logarithmic.
As the interactions are fully characterized by a dimensionless potential $V$, the counterterms,
$V_{ct}$, will be dimensionless as well. This immediately implies that, as long as we are
only interested in the UV divergent part of the diagrams, the interaction vertices in
eq.~(\ref{bgint}) (which are in fact functions of the background fields) can effectively be
treated as constants. A corrolary on this is that vertices with
only holomorphic (or only anti-holomorphic) indices will never contribute to the UV divergences.
Indeed, consider e.g. the diagram in figure (\ref{fig:vertex}).
It gives rise to a contribution of the form,
\begin{eqnarray}
\frac 1 2 \int d \tau _3 d^2 \theta _3 V_{ \gamma \delta  } \ID{}^{ \gamma \bar \alpha } (3-1)
\ID{}^{ \delta \bar \beta } (3-2),
\end{eqnarray}
which, upon partially integrating a fermionic derivative in one of the propagators,
can be seen to only contribute
to the UV finite part of the diagram. Exactly the same reasoning can be made with the effective
propagators appearing later on. The previous also shows that the loop expansion is an expansion
in the number of derivatives. Indeed the $n$-loop contribution to the $ \beta $-functions will %%@
give rise
to $2n-2$ derivatives acting on the fieldstrenghts.

\begin{figure}[h]
\begin{center}
\psfig{figure=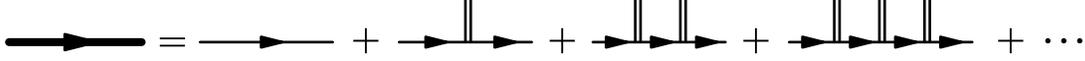}%,height=0.3in}
\caption{The diagrammatic expansion of the effective propagator.
\label{fig:prop}}
\end{center}
\end{figure}

Using the previous observations, one calculates the
relevant effective tree level propagator. It is diagramatically shown in figure %%@
(\ref{fig:prop}).

\noindent Both even and odd numbers of vertices contribute and the final expression reads as,
\begin{eqnarray}
{\cal D}^{ \alpha \bar \beta }(1-2)&=& \ID_+^ { \alpha \bar \beta }(1-2)
+\ID_-^ { \alpha \bar \beta }(1-2),\label{propeff}
\end{eqnarray}
with,
\begin{eqnarray}
\ID_\pm^ { \alpha \bar \beta }(1-2)&=& h_\pm^{ \alpha \bar \beta
}\ID_\pm(1-2)= h_\pm^{ \alpha \bar \beta } D_2 \bar D_2\left( \delta
^{(2)}( \theta _1- \theta _2) \Delta_\pm (1-2)\right), \label{Dpm}
\end{eqnarray}
where $h_\pm^{ \alpha \bar \beta }$ are the inverses of
$h^\pm_{\alpha \bar \beta }\equiv g_{\alpha \bar \beta }\pm F_{\alpha \bar \beta }$,
see eqs.~(\ref{h+-}) and (\ref{h+-inverse}), and
\begin{eqnarray}
\Delta _\pm( \tau )= \frac{1}{2 \pi }\int_m^M \frac{dp}{p}\, e^{\mp
i\,p\, \tau }. \label{t-space prop}
\end{eqnarray}
Note that $ \Delta _+( - \tau )= \Delta _-( \tau )$.

\section{One and two loop contributions}
As a warming up exercise, we calculate the one and two loop contributions in some detail.
\begin{figure}[h]
\begin{center}
\psfig{figure=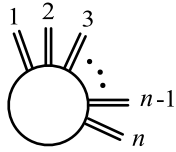}
\caption{A one loop diagram with $n$ vertices. The propagators are the free ones,
eq.~(\ref{propone}).
\label{fig:1loop}}
\end{center}
\end{figure}

\noindent Performing the D-algebra, one finds that a one loop diagram with
$2n$ vertices is given by,
\begin{eqnarray}
\frac{1}{2n\,i}\int d \tau _1d^2 \theta _1d \tau _2 d^2 \theta _2\left(F^{2n}\right)_{ \alpha
\bar \beta }g^{ \alpha \bar \beta } \big(\Delta _+( \tau _1- \tau _2) - \Delta _-( \tau _1- \tau %%@
_2)\big)\delta ^{(3)}(1-2),
\end{eqnarray}
where we used eq.~(\ref{fspot}) and introduced the notation,
\begin{eqnarray}
(F^m)_{ \alpha \bar \beta }\equiv F_{ \alpha \bar \gamma _1}g^{ \bar \gamma _1 \delta _1}
 F_{ \delta _1 \bar \gamma _2}g^{ \bar \gamma _2 \delta _2}
 F_{ \delta _2 \bar \gamma _3}
 \cdots g^{\bar \gamma _{m-1} \delta _{m-1}}F_{ \delta _{m-1}\bar \beta }.
\end{eqnarray}
Because $ \Delta _+( \tau _1- \tau _2) - \Delta _-( \tau _1- \tau _2)$ is an odd function, this %%@
vanishes.
Turning to a one loop diagram with $2n+1$ vertices, one finds an
ultra-violet divergence of the form,
\begin{eqnarray}
- i\,  \frac{ \lambda }{2n+1} \int d \tau d^2\, \theta \,g^{ \alpha
\bar \beta }\left(F^{2n+1}\right)_{ \alpha \bar \beta },
\end{eqnarray}
where,
\begin{eqnarray}
\lambda \equiv \frac{1}{\pi}\ln \left( \frac{M}{m}\right).
\end{eqnarray}
When summing over all loops, we get the total divergent contribution
at one loop,
\begin{eqnarray}
- i\, \lambda \,   \int d \tau d^2 \theta\,  g^{ \alpha \bar \beta }
\left(\mbox{arcth}\,F\right)_{ \alpha \bar \beta },\label{totdiv1}
\end{eqnarray}
where,
\begin{eqnarray}
\left(\mbox{arcth}F\right)_{ \alpha  \bar \beta  }\equiv F_{ \alpha \bar \beta }+\frac 1 3 F_{ %%@
\alpha \bar \gamma }\,g^{ \bar \gamma \delta }\,F_{ \delta \bar \epsilon }\,g^{ \bar \epsilon %%@
\eta}\,F_{\eta \bar \beta }+\cdots
\end{eqnarray}
Using minimal subtraction, this gives us the bare potential through order
$\hbar$,
\begin{eqnarray}
V_{bare}=V+ i\, \lambda   \, g^{ \alpha \bar \beta }
\left(\mbox{arcth}\,F\right)_{ \alpha \bar \beta }.
\end{eqnarray}

Let us first discuss the UV properties of the model at one loop. One reads
from eq.~(\ref{totdiv1}) that the non-linear $ \sigma $-model defined by
eqs.~(\ref{nlsm1}) and (\ref{nlsm2}) is UV finite at one loop provided,
\begin{eqnarray}
g^{ \alpha \bar \beta }\left(\mbox{arcth}\,F\right)_{ \alpha \bar \beta }=0,\label{duydef}
\end{eqnarray}
holds. This equation appeared elsewhere as well. Indeed in \cite{Corrigan:1982th},
the well known four dimensional instanton equations were generalized to higher dimensions.
The generic
class of such configurations are now known as stable holomorphic bundles \cite{duy}. They are
easily characterized if we limit ourselves to a flat (even dimensional and Euclidean)
spacetime. Passing to complex coordinates, one finds using the Bianchi identities that magnetic
fields which satisfy the linear relations,
\begin{eqnarray}
&&F_{ \alpha \beta }=F_{ \bar \alpha \bar \beta }=0, \nonumber\\
&&g^{ \alpha \bar \beta }F_{ \alpha \bar \beta }=0,\label{duys}
\end{eqnarray}
automatically solve the Yang-Mills equations of motion. The first line is the holomorphicity
condition, the second is the stability condition. Later on, these equations were (at least for
sufficiently low dimensions) recognized as BPS equations for supersymmetric Yang-Mills theories
(for a detailed discussion, see e.g.~\cite{koerber:thesis} and references therein).
The starting point of \cite{lies} was the most general deformation of the Maxwell action
involving higher powers of the fieldstrength but excluding derivatives acting on the
fieldstrengths. Requiring that solutions of the type given in eq.~(\ref{duys}) still exist,
uniquely fixes the deformation of the Maxwell action: it is the Born-Infeld action. The
holomorphicity condition in eq.~(\ref{duys}) remains unchanged while the stability condition
gets deformed to precisely eq.~(\ref{duydef}). Concluding, we see that requiring the model
defined by eqs.~(\ref{nlsm1}) and (\ref{nlsm2}) to be UV finite (at one loop) can be
reinterpreted as requiring that the holomorphic bundle satisfies the deformed stability
condition as well.

Turning back to the $\beta $-functions, we find using eq.~(\ref{bfundef}), that they vanish
provided,
\begin{eqnarray}
g^{ \beta \bar \gamma } \partial _ \alpha \left(\mbox{arcth}F
\right)_{ \beta \bar \gamma }=0,\label{loeom}
\end{eqnarray}
holds. Eq.~(\ref{loeom}) arises as the equation of motion for the Born-Infeld action. Indeed,
varying the Born-Infeld action $ {\cal S}_{BI}$,
\begin{eqnarray}
{\cal S}_{BI}=-\int d{}^{2m}X\,\sqrt{\det h_+},
\end{eqnarray}
yields,
\begin{eqnarray}
\delta {\cal S}_{BI}=\int d{}^{2m}X\,\sqrt{\det h_+}\, \delta A_a {\cal G}^{ab} {\cal G}^{cd}
\partial _c F_{db},\label{eom7}
\end{eqnarray}
where $ {\cal G}^{ab}$ is defined in eq.~(\ref{calG}).
Passing to complex coordinates and implementing the holomorphicity, $F_{ \alpha \beta }=F_{ \bar %%@
\alpha \bar \beta }=0$, we get from eq.~(\ref{eom7}) the equations of motion,
\begin{eqnarray}
{\cal G}^{ \beta \bar \gamma } \partial_ \beta  F_{ \bar \gamma  \alpha }=0,
\end{eqnarray}
which upon using the Bianchi identities reduces to eq.~(\ref{loeom}).
\begin{figure}[h]
\begin{center}
\psfig{figure=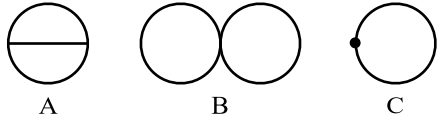}%,height=0.3in}
\caption{The two 2-loop diagrams and the contribution arising from the one loop counter term.
The effective propagator, eq.~(\ref{propeff}), is used.
\label{fig:2loop}}
\end{center}
\end{figure}

We now turn to the two loop contributions. Technical details of the calculation can be
found in appendix B. Diagram A in fig. (\ref{fig:2loop}) results in,
\begin{eqnarray}
-\frac{ \lambda  ^2}{2}\int d \tau d^2 \theta \,V_{ \bar \beta _1
\alpha _2 \alpha _3} V_{ \alpha _1 \bar \beta _2 \bar \beta
_3}\left( {\cal G}^{ \alpha_1 \bar \beta _1} {\cal G}^{ \alpha _2
\bar \beta _2} {\cal B}^{ \alpha _3 \bar \beta _3}+ {\cal B}^{
\alpha_1 \bar \beta _1} {\cal G}^{ \alpha _2 \bar \beta _2} {\cal
G}^{ \alpha _3 \bar \beta _3} \right),\label{2loop:diagrA}
\end{eqnarray}
where,
\begin{eqnarray}
{\cal B}^{ab}\equiv \frac{1}{2\,i}\big(h_+^{ab}-h_-^{ab}\big).
\end{eqnarray}
Diagram B gives,
\begin{eqnarray}
\frac{ \lambda  ^2}{2}\int d \tau d^2 \theta \, {\cal G}^{ \alpha
\bar \beta } {\cal G}^{ \gamma \bar \delta } V_{ \alpha \bar \beta
\gamma \bar \delta }. \label{2loop:diagrB}
\end{eqnarray}
Adding the two gives,
\begin{eqnarray}
-i\frac{ \lambda  ^2}{2} \int d \tau d^2 \theta \,{\cal G}^{ \alpha
\bar \beta } \partial _ \alpha
\partial _{ \bar \beta }\left(g^{ \gamma \bar \delta }
\left(\mbox{arcth}\,F\right)_{ \gamma \bar \delta  }\right).
\label{2loop:A+B}
\end{eqnarray}
Finally, the subdivergence (diagram C), which arises from the one
loop counter term, has the same structure and it is given by,
\begin{eqnarray}
 i\;\lambda^2\int d \tau d^2 \theta \, {\cal G}^{ \alpha \bar \beta }
\partial _ \alpha
\partial _{ \bar \beta }\left(g^{ \gamma \bar \delta }
\left(\mbox{arcth}\,F\right)_{ \gamma \bar \delta  }\right).
\end{eqnarray}
Adding these contributions gives us the bare potential through two loops.
\begin{eqnarray}
V_{bare}=V+ \lambda\, V_{(1)}+ \lambda ^2\, V_{(2)},\label{Vbare2}
\end{eqnarray}
where
\begin{eqnarray}
V_{(1)}&=&i    g^{ \alpha \bar \beta }
\left(\mbox{arcth}\,F\right)_{ \alpha \bar \beta } \nonumber\\
V_{(2)}&=& - \frac{i}{2}\, {\cal G}^{ \alpha \bar \beta } \partial _
\alpha \partial _{ \bar \beta }\left(g^{ \gamma \bar \delta }
\left(\mbox{arcth}\,F\right)_{ \gamma \bar \delta
}\right) \label{2loop:result}\\
&=& - \frac 1 2 \,
\psfig{figure=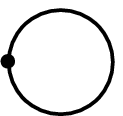,bbllx=0,bblly=13,bburx=33,bbury=31}
\, S\;. \nonumber
\end{eqnarray}

\vspace{.1cm}

\noindent In the last line we used the diagrammatic notation
introduced in appendix \ref{app conv}. One easily verifies that the
two loop ${\cal O}( \lambda ^2)$ counterterm agrees with the
renormalization group result, eq.~(\ref{rg}).

As the bare potential does not get an order $ \lambda  $ correction at two loops, the
$ \beta $-function will not be modified at this order and as a consequence, the Born-Infeld %%@
action
receives no two derivative corrections. This result agrees with \cite{andreevtseytlin}.

\section{Three loop result}

\begin{figure}[h]
\begin{center}
\psfig{figure=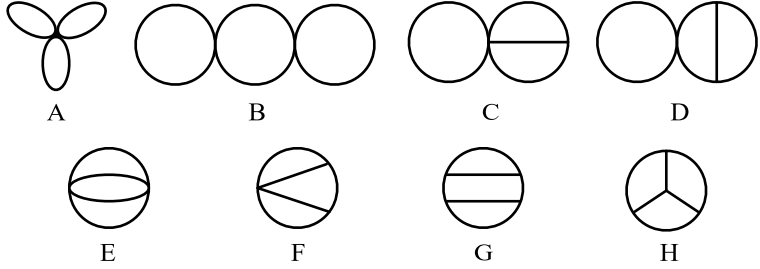}%,height=0.3in}
\caption{All the 3-loop diagrams.
The effective propagator, eq.~(\ref{propeff}), is used.
\label{fig:3loop}}
\end{center}
\end{figure}

The 3-loop diagrams are shown in fig. (\ref{fig:3loop}). As is
explained in more detail in appendix \ref{app 3l}, the
three loop diagrams give rise to terms proportional to $\lambda^3$
as well as terms linear in $\lambda$. It is a non-trivial check on
the calculation that, with the contributions from the one and two
loop counterterms taken into account, there are no terms quadratic
in $\lambda$ present at this order. This must be the case, since
there was no contribution to the $\beta$-function at the two loop
level. The $\lambda^3$ terms can be expressed quite concisely as
\begin{eqnarray}
\begin{split}
V_{(3)}&= - \frac{1}{3}\, {\cal G}^{ \alpha \bar \beta }
\partial _ \alpha \partial _{ \bar \beta }V_{(2)} -\frac{i}{12}
\left( h_+^{\alpha \bar \beta}h_+^{\gamma \bar \delta} - h_-^{\alpha
\bar \beta}h_-^{\gamma \bar \delta}\right)
\partial_\alpha \partial_{\bar \delta} V_{(1)}
\partial_\gamma \partial_{\bar \beta} V_{(1)}\\
&= - \frac{1}{3}\,
\psfig{figure=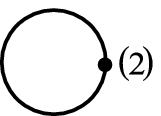,bbllx=0,bblly=12,bburx=44,bbury=31}
- \frac{i}{12} \left(
\psfig{figure=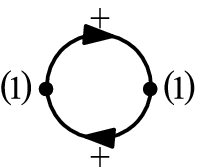,bbllx=0,bblly=20,bburx=56,bbury=46}
\, \, - \, \,
\psfig{figure=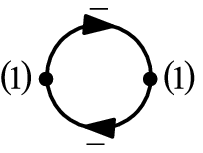,bbllx=0,bblly=20,bburx=56,bbury=46}
\right) \, .\label{3loop:RG}
\end{split}
\end{eqnarray}
A very strong check on the calculation is the fact that this
precisely agrees with the renormalization group equations in
eq.~(\ref{rg})! In the second line we again used the diagrammatic
notation explained in appendix (\ref{app conv}). Adding the terms
linear in $\lambda$ to the one loop result, we find up to terms
containing four derivatives of the fieldstrength
\begin{eqnarray}
V_{(1)}&=&i    g^{ \alpha \bar \beta } \left(\mbox{arcth}\,\tilde
F\right)_{ \alpha \bar \beta } - \frac{1}{48} S_{ab\alpha \bar
\beta} S_{cd\gamma \bar \delta}\; h_+^{bc}h_+^{da}\left( {\cal
G}^{\alpha \bar \delta} {\cal B}^{\gamma \bar \beta} + {\cal
B}^{\alpha \bar \delta} {\cal G}^{\gamma \bar \beta} \right) + {\cal
K}V_{(1,1)}\;.\label{result}
\end{eqnarray}
Here $\tilde F$ is the fieldstrength associated with the gauge
potential $\tilde A$, related to the original $A$ by the field
redefinition
\begin{eqnarray}
\begin{split}
\tilde A_{\alpha} &= A_{\alpha} + \frac{1}{24}\partial_{\alpha}
\left( F_{\beta_1 \bar \gamma_2,\bar \gamma_3} F_{\beta_2 \bar
\gamma_1,\beta_3} h_+^{\beta_1 \bar \gamma_1}
h_+^{\beta_2 \bar \gamma_2} {\cal G}^{\beta_3 \bar \gamma_3} \right)\\
&= A_{\alpha} + \frac{1}{24}\partial_{\alpha}\,
\psfig{figure=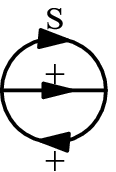,bbllx=0,bblly=21,bburx=31,bbury=48}\,;
\\ \\
\tilde A_{\bar \alpha} &= A_{\bar \alpha} +
\frac{1}{24}\partial_{\bar \alpha} \left( F_{\beta_1 \bar
\gamma_2,\bar \gamma_3} F_{\beta_2 \bar \gamma_1,\beta_3}
h_-^{\beta_1 \bar \gamma_1} h_-^{\beta_2 \bar \gamma_2} {\cal
G}^{\beta_3 \bar \gamma_3} \right)\\
&= A_{\bar \alpha} + \frac{1}{24}\partial_{\bar \alpha}\,
\psfig{figure=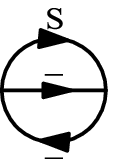,bbllx=0,bblly=21,bburx=31,bbury=48}\,.
\label{field redef}
\end{split}
\end{eqnarray}

\vspace{.3cm}

\noindent Note that we can omit the tilde on the fields appearing in the
second term in eq.~(\ref{result}), because this has only an effect on the terms
which are higher order in $\hbar$. We also introduced (see \cite{wyllard})
\begin{eqnarray}
S_{abcd}=\partial_a \partial_b F_{cd} +
h_+^{ef}\, \partial_a
F_{c e}\, \partial_b F_{ df}
-h_+^{ef}\, \partial_a
F_{d e}\, \partial_b F_{ cf}.
\label{Scurv}
\end{eqnarray}
The use of Latin indices indicates a summation over real
coordinates, where the use of Greek indices, as usual, refers to a
summation over complexified holomorphic or anti-holomorphic
coordinates. Finally, in the last term of eq.~(\ref{result}), ${\cal
K}$ is a derivative operator acting on the one loop part of
$V_{(1)}$ (see eq.~(\ref{loopexp}) for notation). This term can be
written in a fairly transparent way by again making use of our
diagrammatic notation:
\begin{eqnarray}
{\cal K}V_{(1,1)} &=& \frac{1}{24}  \left(
\psfig{figure=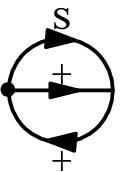,bbllx=0,bblly=21,bburx=33,bbury=48} \right. \, \, - \, \,
\psfig{figure=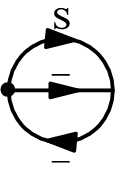,bbllx=0,bblly=21,bburx=33,bbury=48} \, \, + \, \,
\psfig{figure=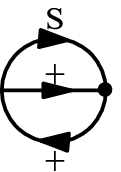,bbllx=0,bblly=21,bburx=33,bbury=48} \, \, - \, \,
\psfig{figure=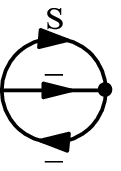,bbllx=0,bblly=21,bburx=33,bbury=48}\nonumber \\
&& \quad  \quad \, \,
\psfig{figure=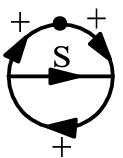,bbllx=0,bblly=21,bburx=33,bbury=48}  \, \, + \, \,
\psfig{figure=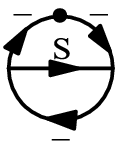,bbllx=0,bblly=21,bburx=33,bbury=48} \, \, + \, \,
\psfig{figure=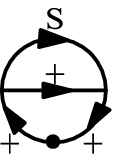,bbllx=0,bblly=21,bburx=33,bbury=48} \, \, + \, \,  \left.
\psfig{figure=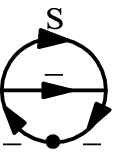,bbllx=0,bblly=21,bburx=33,bbury=48} \right) \\
&+& \frac{1}{48}  \left(
\psfig{figure=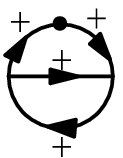,bbllx=0,bblly=21,bburx=33,bbury=48} \right.  \, \, - \, \,
\psfig{figure=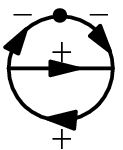,bbllx=0,bblly=21,bburx=33,bbury=48} \, \, - \, \,
\psfig{figure=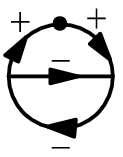,bbllx=0,bblly=21,bburx=33,bbury=48} \, \, + \, \,  \left.
\psfig{figure=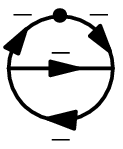,bbllx=0,bblly=21,bburx=33,bbury=48} \right)\nonumber
\end{eqnarray}
Since up to this order, we can put $V_{(1,1)}$ to zero and, in general, a field
redefinition has no physical consequences, we arrive at the following correction to the
stability condition for holomorphic
vector bundles,
\begin{eqnarray}
g^{ \alpha \bar \beta } \left(\mbox{arcth}\, F\right)_{ \alpha \bar
\beta } + \frac{i}{48} S_{ab\alpha \bar \beta} S_{cd\gamma \bar
\delta}\; h_+^{bc}h_+^{da}\left( {\cal G}^{\alpha \bar \delta} {\cal
B}^{\gamma \bar \beta} + {\cal B}^{\alpha \bar \delta} {\cal
G}^{\gamma \bar \beta} \right) =0.
\end{eqnarray}
This is exactly the stability condition which follows from the action
presented in \cite{wyllard} (see \cite{koerber:thesis}):
\begin{eqnarray}
{\cal S} = - \tau_9 \int d^{10}x\; \sqrt{h_+}\bigg[ 1 &+&
\frac{1}{96} \Big( \frac 1 2 h_+^{ab}h_+^{cd}
S_{bc} S_{d a}\nonumber\\
&-&
h_+^{c_2a_1}h_+^{a_2 c_1}h_+^{d_2 b_1}h_+^{b_2 d_1}
S_{a_1 a_2 b_1 b_2}S_{c_1 c_2 d_1 d_2}\Big)\bigg],
\end{eqnarray}
where
\begin{eqnarray}
S_{ab} = h_+^{cd} S_{abcd}.
\end{eqnarray}

\section{Conclusions and outlook}

In the present paper we calculated the $ \beta $-functions through three loops for an open
string $ \sigma $-model in the presence of $U(1)$ background. Requiring them to vanish is then
reinterpreted as the string equations of motion for the background. Upon integration this yields
the low energy effective action. Doing the calculation in $N=2$ boundary superspace
significantly simplified the calculation. The one loop contribution gives the effective action
to all orders in $ \alpha {}'$ in the limit of a constant fieldstrength. The result is the well
known Born-Infeld action. The absence of a two loop contribution to the $ \beta $-function shows
the absence of two derivative terms in the action. Finally the three loop contribution gives the
four derivative terms in the effective action to all orders in $ \alpha {}'$. Modulo a
field redefinition we find complete agreement with the proposal made in \cite{wyllard}.

By doing the calculation in $N=2$ superspace, we get a nice geometric characterization of
UV finiteness of the non-linear $ \sigma $-model: UV finiteness is guaranteed provided that
the background is a deformed stable holomorphic bundle.

\begin{figure}[h]
\begin{center}
\psfig{figure=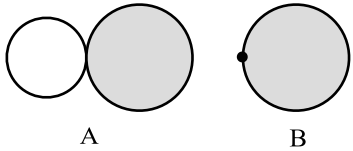}%,height=0.3in}
\caption{A general diagram with external loop (A) and that same
diagram with the external loop replaced by the one loop counterterm
(B). \label{fig:blob}}
\end{center}
\end{figure}

An immediate question is whether the present program can be pushed to higher orders.
Of course, already at four loops this procedure becomes extremely
cumbersome. There might however be general arguments that lead to
considerable simplifications. First of all, consider a diagram with
an external loop, which will in general look like diagram A of
figure (\ref{fig:blob}), where the bigger circle with shaded area
can be any diagram\footnote{For some examples, see diagrams A - D of
figure (\ref{fig:3loop})}. It is not very difficult to show that
diagram B of figure (\ref{fig:blob}), resulting from replacing that
external loop with the one loop counterterm, will contain a term
which exactly cancels the original diagram. As a consequence,
diagrams with (one or more) external loops cannot contribute to the
$\beta$-function. This is of course part of the bigger
renormalization group picture. Diagrams which factorize will never
contribute to the $\beta$-function, because the divergences
encountered in the corresponding loop-integrals are already
accounted for at a lower-loop level.

\begin{figure}[h]
\begin{center}
\psfig{figure=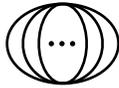}%,height=0.3in}
\caption{An $n$-loop ``onion'' diagram. \label{fig:najuin}}
\end{center}
\end{figure}

More important simplifications might arise from an interesting
observation made in \cite{wyllard}. There Niclas Wyllard noted that
$S_{abcd}$ which was introduced in eq.~(\ref{Scurv}) can be viewed
as the curvature tensor for a non-symmetric connection. Once this is
better understood, this could lead to a method giving results to all
order in the derivatives. Indeed the leading contribution to the $
\beta $-functions comes from the $n$-loop ``onion'' diagram shown in
fig. ~(\ref{fig:najuin}) which can be explicitly calculated in a
reasonably straightforward way. The remainder of the $ \beta
$-function should then follow as some sort of covariantization. This
point of view is presently under investigation \cite{wip}.

Finally, a natural question which arises here is whether the present
method extends to the non-abelian case. In that case the coupling to
the gauge fields involves the introduction of a Wilson line. The
path-ordering can be undone through the introduction of auxiliary
fields, \cite{dorn}, and a first exploration was performed in
\cite{klaus}. However, before the present analysis can be done for
the non-abelian case, one needs to extend the superspace formulation
in \cite{Koerber:2003ef} such as to include Wilson loops and the
auxiliary formulation of \cite{dorn}. This would certainly lead to
significant information on non-abelian deformed stable holomorphic
bundles. We leave this interesting question to future investigation.

\acknowledgments

\bigskip

We thank Klaus Behrndt, Marc Grisaru, Chris Hull and Paul Koerber for useful discussions.
All authors are supported in part by the Belgian Federal Science Policy Office
through the Interuniversity Attraction Pole P5/27 and in part by the European
Commission FP6 RTN programme MRTN-CT-2004-005104. SN, AS and AW are supported in part by the
``FWO-Vlaanderen'' through project G.0428.06, and WT through  project G.0235.05.

\appendix

\section{Conventions, notations and identities}\label{app conv}

Both the worldsheet as well as the
target space carry a flat euclidean metric. The target space has dimension $d=2n$, $n\in\IN$ and %%@
we use roman indices to denote real
and greek indices for complex coordinates. We write the metric as $g_{ab}$ and we have that %%@
$g_{ab}= \delta _{ab}$. The magnetic fields,
$F_{ab}$, are incorporated in the open string metric(s), $h_{ab}^\pm$,
\begin{eqnarray}
h^\pm_{ab}\equiv g_{ab}\pm F_{ab},\label{h+-}
\end{eqnarray}
and we obviously have that $h_{ab}^+=h_{ba}^-$.
The inverse, $h_\pm^{ab}$, is defined by,
\begin{eqnarray}
h^{ac}_+\,h^+_{cb}=h^{ac}_-\,h^-_{cb}= \delta ^a_b.\label{h+-inverse}
\end{eqnarray}
In addition we define,
\begin{eqnarray}
{\cal G}^{ab}\equiv \frac 1 2 \left(h_+^{ab}+h_-^{ab}\right).\label{calG}
\end{eqnarray}
In complex coordinates, the metric of the target space
is given by,
\begin{eqnarray}
g_{\alpha \beta}=g_{\bar\alpha
\bar\beta}=0, \qquad g_{\alpha \bar\beta}=\frac 1 2 \,\delta_{\alpha \bar\beta}.
\end{eqnarray}
The $N=2$ supersymmetry requires the $U(1)$ bundle to be holomorphic,
\begin{eqnarray}
F_{ \alpha \beta }= F_{ \bar \alpha \bar \beta }=0,
\end{eqnarray}
which implies that,
\begin{eqnarray}
h_{ \alpha \beta }^\pm= h_{ \bar \alpha \bar \beta }^\pm=0.
\end{eqnarray}

For much of the superspace techniques, we refer to the ``bible'': \cite{Gates:1983nr}!
The $N=2$ boundary superspace is parameterized by two bosonic coordinates $ \tau\in\IR $,
and $ \sigma \in\IR$, $ \sigma \geq 0$, and two fermionic coordinates $ \theta $ and $
\bar \theta $. The fermionic derivatives are defined by,
\begin{eqnarray}
D \theta = \bar D \bar \theta =1,\qquad D \bar \theta = \bar D \theta =0,
\end{eqnarray}
and,
\begin{eqnarray}
D^2= \bar D^2=0,\qquad \{D,\bar D\}= \partial _ \tau .
\end{eqnarray}
The superspace integration measure is defined by,
\begin{eqnarray}
\int d^ 2 \theta \,\bar \theta \theta =1.
\end{eqnarray}
Some definitions involving $ \delta $-functions,
\begin{eqnarray}
\delta ^{(2)}( \theta _1- \theta _2)&\equiv& ( \bar \theta _1-\bar \theta _2)
( \theta _1- \theta _2), \nonumber\\
\delta ^{(3)}(1-2)&\equiv& \delta ( \tau _1- \tau _2) \delta ^{(2)}( \theta _1 - \theta _2),
\nonumber\\
\delta ^{(4)}(1-2)&\equiv& \delta^{(2)}
( \sigma  _1- \sigma  _2) \delta ^{(2)}( \theta _1 - \theta _2),
\end{eqnarray}
and some useful identities,
\begin{eqnarray}
&&\delta ^{(2)}( \theta _1- \theta _2) \delta ^{(2)} ( \theta _2- \theta _1)=0, \nonumber\\
&&\delta ^{(2)}( \theta _1- \theta _2) D_a\left(\delta ^{(2)} ( \theta _2- \theta _1)
f( \tau _2- \tau _1)\right)=0, \nonumber\\
&&\delta ^{(2)}( \theta _1- \theta _2)\bar D_a\left( \delta ^{(2)} ( \theta _2- \theta _1)
f( \tau _2- \tau _1)\right)=0, \nonumber\\
&&\delta ^{(2)}( \theta _1- \theta _2) \left(D_a \bar D_a\delta ^{(2)} ( \theta _2- \theta _1)
f( \tau _2- \tau _1)\right)= \nonumber\\
&&-\delta ^{(2)}( \theta _1- \theta _2)\left(\bar D_a  D_a\delta ^{(2)} ( \theta _2- \theta _1)
f( \tau _2- \tau _1)\right)=
\delta ^{(2)}( \theta _1- \theta _2)f( \tau _2- \tau _1),
\nonumber\\
\end{eqnarray}
where the subindex $a$ is $1$ or $2$ and where we did not sum over repeated subindices $a$.

In keeping track of all contributions to the bare potential and
checking the renormalization group equations it proved to be
very helpful to introduce a diagrammatic notation for the index
structure of the different contributions. One of the most attractive
features of this notation is that the diagram of the index structure
is, as will become clear, exactly the same as the diagram of the
full contribution it corresponds to. Of course, this same feature is
also a possible cause for confusion. We hope however that the
precise meaning of the diagram will always be clear from the
context. When diagrams are used several times in the same formula,
not necessarily always with the same meaning, the part that
indicates the index structure only (without the momentum integral
and symmetry factor), will appear in between brackets. (For an
example of this, see eq.~(\ref{2loop:diag}) below.) More concretely,
we introduce the notations

\begin{equation}
h_{\pm}^{\alpha\bar{\beta}} =
\psfig{figure=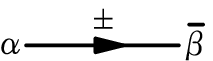,bbllx=0,bblly=4,bburx=59,bbury=16}\,,\label{diag.prop}
\end{equation}

\begin{equation}
{\cal G}^{\alpha \bar \beta} =\frac 1 2 \left(h_+^{ \alpha \bar \beta }+
h_-^{ \alpha \bar \beta } \right)=
\psfig{figure=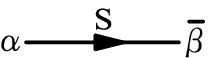,bbllx=0,bblly=4,bburx=59,bbury=16}\,,
\end{equation}

\begin{equation}
\partial_\gamma F_{\alpha \bar \beta} =
\partial_\alpha F_{\gamma \bar \beta} =
\psfig{figure=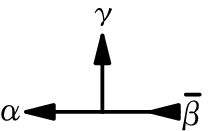,bbllx=0,bblly=15,bburx=58,bbury=37}
\end{equation}
and

\begin{equation}
\partial_{\bar \gamma} F_{\alpha \bar \beta} =
\partial_{\bar \beta} F_{\alpha \bar \gamma} =
\psfig{figure=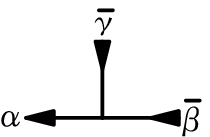,bbllx=0,bblly=15,bburx=58,bbury=37}\,.
\end{equation}

\vspace{.4cm}

\noindent Here it should be clear that
\psfig{figure=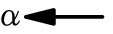,bbllx=0,bblly=3,bburx=30,bbury=11}
and
\psfig{figure=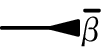,bbllx=0,bblly=3,bburx=29,bbury=11}
represent derivatives on the field strength or potential and should not be
confused with the object defined in eq.~(\ref{diag.prop})
corresponding to the propagator.

It is very easy to compute derivatives of different expressions by
using (for example)
\begin{eqnarray}
\partial_{\bar \gamma} h_\pm^{\alpha \bar \beta} = \pm\, h_\pm^{\alpha \bar
\delta} h_\pm^{\varepsilon \bar \beta}\, \partial_{\bar \gamma}
F_{\varepsilon \bar \delta}.
\end{eqnarray}
In diagrammatic form this becomes
\begin{equation}
\partial_{\bar \gamma}\left( \psfig{figure=hplusmin_59x16.eps,bbllx=0,bblly=4,bburx=59,bbury=16} %%@
\right) =
\pm
\psfig{figure=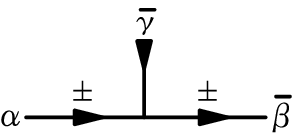,bbllx=0,bblly=14,bburx=84,bbury=36}.
\end{equation}

\vspace{.4cm}

\noindent The same formula evidently also holds for derivatives with
respect to holomorphic coordinates. Also very useful is the
following identity:
\begin{equation}
\partial_{\bar{\beta}} V_{(1,1)}
=\partial_{\bar{\beta}}\left(i\, g^{ \gamma \bar \delta }
\left(\mbox{arcth}\, F\right)_{ \gamma \bar \delta }
 \right)
= i {\cal G}^{\gamma \bar \delta} \partial_{\bar \beta} F_{\gamma \bar \delta}
= i
\psfig{figure=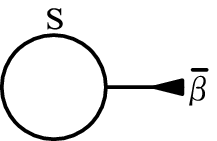,bbllx=0,bblly=15,bburx=60,bbury=39}.
\end{equation}

\vspace{.4cm}

\noindent Applying yet another derivative results in
\begin{equation}
\partial_{\alpha}\partial_{\bar{\beta}} V_{1} = i \left[
\psfig{figure=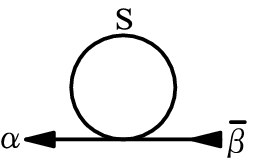,bbllx=0,bblly=22,bburx=71,bbury=47}
\, \, + \frac{1}{2} \Bigg(
\psfig{figure=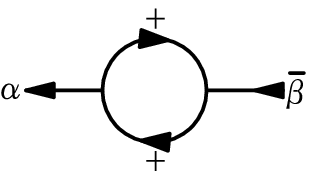,bbllx=0,bblly=22,bburx=88,bbury=47}
\, \, - \, \,
\psfig{figure=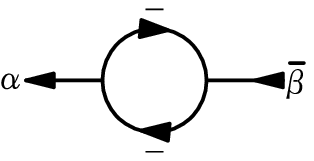,bbllx=0,bblly=22,bburx=88,bbury=47}
\Bigg)\right]\label{ppartialV}
\end{equation}

\vspace{.2cm}

\section{Explicit two loop computation}\label{app 2l}

\begin{figure}[h]
\begin{center}
\psfig{figure=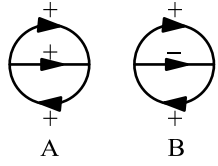} \caption{The basic
non-trivial diagrams at two loops. \label{fig:2loop_basic}}
\end{center}
\end{figure}

As an illustration of how the calculation was done, we will now
explicitly compute the two loop contribution. Let us first consider
diagram A in fig. (\ref{fig:2loop}). Of course the effective
propagator, eq.~(\ref{propeff}), has a certain direction, so that
diagram A actually represents all possible inequivalent topologies
with directed lines. Since the righthand side of eq.~(\ref{propeff})
also consists of two terms, in the end we have to consider all
possible distinct diagrams with directed lines and all possible
combinations of positive and negative `frequency' parts of the
propagator (eq.~(\ref{Dpm})). It turns out that the only momentum
integrals that have to be performed explicitly, are the ones
belonging to the two diagrams in figure (\ref{fig:2loop_basic}). All
other integrals are either zero (or finite, and therefore irrelevant
for our beta function computation), or are related to these
ones by complex conjugation or partial integration.

Let us start with diagram A of figure (\ref{fig:2loop_basic}). The Feynman rules for this
diagram give:
\begin{eqnarray}
\frac 1 2 \left\langle V_{ \bar \beta _1 \alpha _2 \alpha _3}(1) V_{ \alpha _1
\bar \beta _2 \bar \beta _3}(2) \ID_+^{ \alpha_1 \bar \beta _1}(2-1)
\ID_+^{ \alpha _2 \bar \beta _2}(1-2) \ID_+^{ \alpha _3 \bar \beta
_3}(1-2)\right\rangle_{1,2}\;,
\end{eqnarray}
where $\langle \rangle_1$ means integrating over $\tau_1$, $\theta_1$ and $\bar
\theta_1$. Using eq.~(\ref{Dpm}) and performing the D-algebra leads
to
\begin{eqnarray}
\frac 1 2 \left\langle V_{ \bar \beta _1 \alpha _2 \alpha _3} V_{
\alpha _1 \bar \beta _2 \bar \beta _3} h_+^{ \alpha_1 \bar \beta _1}
h_+^{ \alpha _2 \bar \beta _2} h_+^{ \alpha _3 \bar \beta _3}{\cal
I}_A\right\rangle_1\;,
\end{eqnarray}
with
\begin{eqnarray}
{\cal I}_A = \int_{-\infty}^{+\infty} d\tau \;
\Delta_+(\tau)\Delta_+(\tau)
\partial_{\tau}\Delta_-(\tau)\;,\label{I}
\end{eqnarray}
where we made the change of variable $\tau_2 \rightarrow \tau \equiv
\tau_1 - \tau_2$ and made use of $ \Delta _+( - \tau )= \Delta _-(
\tau )$. Inserting the explicit form of the coordinate space
propagators eq.~(\ref{t-space prop}) and using the fact that
\begin{eqnarray}
\partial_\tau \Delta_\pm (\tau) = \frac{1}{2\pi\tau} (e^{\mp iM\tau}- e^{\mp
im\tau})\;,
\end{eqnarray}
we arrive at the following expression:
\begin{eqnarray}
{\cal I}_A &=& \frac{1}{(2\pi)^3}\int_m^M \frac{dp}{p} \int_m^M
\frac{dq}{q} \int \frac{d\tau}{\tau}\;
\left(e^{-i(p+q-M)\tau}-e^{-i(p+q-m)\tau}\right) \\
&=& -\frac{i\pi}{(2\pi)^3}\int_1^\Lambda \frac{dp}{p} \int_1^\Lambda
\frac{dq}{q}\;[\epsilon(p+q-\Lambda)-\epsilon(p+q-1)]\;,\label{I
p-space}
\end{eqnarray}
where we used the sign function $\epsilon(p)=1$ for $p>0$ and
$\epsilon(p)=-1$ for $p<0$. In the last line all momenta are
expressed in units of $m$ and again $\Lambda = M/m$. This integral
can easily be written in terms of logarithms and dilogarithms,
\begin{eqnarray}
{\cal I}_A = \frac{i}{(2\pi)^2}\left[ \ln(\Lambda-1)\ln\Lambda +
\Li_2\left(\frac{1}{\Lambda}\right)-
\Li_2\left(\frac{\Lambda-1}{\Lambda}\right)\right]\;,\label{finite
L}
\end{eqnarray}
with (the restriction to $\vert x \vert \leq 1$ is only required for
the second equality)
\begin{eqnarray}
\Li_2(x) = -\int_0^x dz\; \frac{\ln(1-z)}{z} = \sum_{k=1}^\infty
\frac{x^k}{k^2}\;, \quad \vert x \vert \leq 1\;.\label{poly2}
\end{eqnarray}
As we are only interested in the UV divergent part
of ${\cal I}_A$, we only need the behavior of eq.
(\ref{finite L}) for $\Lambda >> 1$:\;\footnote{The first term on
the right hand side is irrelevant when we only consider two loop contributions.
However, at the three loop level (for instance for diagrams C and D
of fig. (\ref{fig:3loop})) this result will be multiplied by
$\log(\Lambda)$, coming from the extra loop. This means that the
first term will come into play, while the subleading terms of
order $1/\Lambda$ and higher will continue to be irrelevant for the
UV behavior.}
\begin{eqnarray}
{\cal I}_A = -\frac{i}{24}+\frac{i}{(2\pi)^2}\ln^2(\Lambda) + {\cal
O}\left( \frac{1}{\Lambda}\right) \;,
\end{eqnarray}
where we used $\Li_2(1)=\zeta(2) = \pi^2/6$, as is clear from eq.
(\ref{poly2}). Putting everything together, we find the following
contribution from diagram A of figure (\ref{fig:2loop_basic}) to eq.
(\ref{2loop:diagrA}):
\begin{eqnarray}
i\;\frac{\lambda ^2}{8}\int d \tau d^2 \theta \, V_{ \bar \beta _1
\alpha _2 \alpha _3} V_{ \alpha _1 \bar \beta _2 \bar \beta _3}
h_+^{ \alpha_1 \bar \beta _1} h_+^{ \alpha _2 \bar \beta _2} h_+^{
\alpha _3 \bar \beta _3}\;.
\end{eqnarray}
It is readily verified that this is indeed one of the terms
appearing in eq.~(\ref{2loop:diagrA}).

Turning now to diagram B of
figure (\ref{fig:2loop_basic}), one finds that the equivalent of ${\cal I}_A$ for this
diagram is
\begin{eqnarray}
{\cal I}_B = \int_{-\infty}^{+\infty} d\tau \;
\Delta_+(\tau)\Delta_-(\tau)
\partial_{\tau}\Delta_-(\tau)\;,
\end{eqnarray}
However, this integral at its turn can be related to ${\cal I}_A$ by
partial integration and complex conjugation:
\begin{eqnarray}
{\cal I}_B = -\frac 1 2 \int_{-\infty}^{+\infty} d\tau \;
\Delta_-(\tau)\Delta_-(\tau)
\partial_{\tau}\Delta_+(\tau) = -\frac1 2 {\cal I}_A^* = \frac1 2 {\cal
I}_A\;,
\end{eqnarray}
where in the last equality we used the fact that ${\cal I}_A$ is purely
imaginary. Diagram B of figure (\ref{fig:2loop}) almost trivially leads to
(\ref{2loop:diagrB}), so that it turns out that diagram A of figure
(\ref{fig:2loop_basic}) gives the only nontrivial integral we have to compute at two
loops. The total contribution from diagrams A and B in figure
(\ref{fig:2loop}) can very nicely and suggestively be written using
the diagrammatic notation for the index structure explained in the
previous section:
\begin{equation}
\psfig{figure=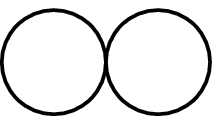,bbllx=0,bblly=13,bburx=61,bbury=31} \, \, + \, \,
\psfig{figure=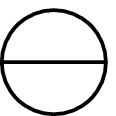,bbllx=0,bblly=13,bburx=31,bbury=31}  = - \frac i 2 %%@
\lambda^2
\left\langle  \psfig{figure=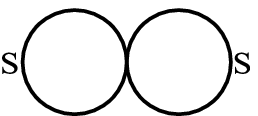,bbllx=0,bblly=13,bburx=72,bbury=31} \, \, + %%@
\frac 1 2
\Bigg(  \psfig{figure=2_loop_setting_sun_spp_31x48.eps,bbllx=0,bblly=21,bburx=31,bbury=48} \, \, %%@
- \, \,
\psfig{figure=2_loop_setting_sun_smm_31x48.eps,bbllx=0,bblly=21,bburx=31,bbury=48}  \Bigg)  %%@
\right\rangle
\label{2loop:diag}
\end{equation}
Finally, the contribution from diagram C in figure (\ref{fig:2loop})
has to be taken into account. Using eq.~(\ref{ppartialV}) one finds
\begin{eqnarray}
\psfig{figure=2_loop_counterterm_33x31.eps,bbllx=0,bblly=13,bburx=33,bbury=31} &=& \lambda^2
\left\langle {\cal G}^{\alpha \bar \beta} \partial_{\alpha}\partial_{\bar{\beta}} V_{1} %%@
\right\rangle \nonumber\\
&=& i \lambda^2
\left\langle  \psfig{figure=2_loop_bril_SS_72x31.eps,bbllx=0,bblly=13,bburx=72,bbury=31} \, \, + %%@
\frac 1 2 \Bigg(  %%@
\psfig{figure=2_loop_setting_sun_spp_31x48.eps,bbllx=0,bblly=21,bburx=31,bbury=48} \, \, - \, \,  %%@
\psfig{figure=2_loop_setting_sun_smm_31x48.eps,bbllx=0,bblly=21,bburx=31,bbury=48}  \Bigg)  %%@
\right\rangle
\end{eqnarray}
Comparing this to (\ref{2loop:diag}) it is easy to understand why
indeed we could write eq.~(\ref{2loop:A+B}) the way we did. Adding
all two loop contributions, we indeed find (\ref{2loop:result}).

While the use of this diagrammatic notation might seem a bit
overdoing it at two loops, it becomes unavoidable at three loops. Of
course, even when using our diagrammatic notation, calculations will
become quite lengthy at that stage, so we will not show any of these
here explicitly. Let us only state that heavy use of this notation
was made in checking eq.~(\ref{3loop:RG}) and working out the
consequences of the field redefinition (\ref{field redef}).

\section{Outline of the three loop calculation}\label{app 3l}

Before we start explaining the general procedure we used for
calculating the three loop contributions, it is useful to understand
some general features for any number of loops. Let $v$ be the number
of vertices in the diagram under consideration, $e$ the number of
edges (propagators) and $l$ the number of loops. These have to
satisfy the topological relation $v-e+l=1$. It follows from the
D-algebra that the number of derivatives appearing in the equivalent
of eq.~(\ref{I}), which we will call $d$, equals
$v-1$.\footnote{This will always equal the number of $\tau$
integrations (number of vertices minus the global position of the
diagram), so that the procedure will always make sense.} From this
we can conclude that the number of remaining momentum integrations
will be $e-d = e-v+1 = l$, which, of course, makes very good sense.
More importantly, the fact that in general there will be more than
one derivative appearing in the equivalent of eq.~(\ref{I}), will
lead to products of multiple sign functions in the equivalent of eq.
(\ref{I p-space}), $d=v-1$ of them to be precise.

Now, let us focus on $l=3$. One always ends up with having to do
three momentum integrals of the type appearing in eq.~(\ref{I
p-space}) with, in general, a product of one, two or three sign
functions. To this end, one separates the 3-dimensional domain of
integration into smaller parts, such that on each part the product
of the appearing sign functions has a definite value. From the
definition of the dilogarithm, eq.~(\ref{poly2}), and the
trilogarithm,
\begin{eqnarray}
\Li_3(x) = \int_0^x dz\; \frac{\Li_2(z)}{z} = \sum_{k=1}^\infty
\frac{x^k}{k^3}\;, \quad \vert x \vert \leq 1\;,\label{poly3}
\end{eqnarray}
it is clear that one ends up with expressions involving
logarithms, dilogarithms and trilogarithms. Fortunately, since we
are only interested in the behavior of these expressions for
$\Lambda >> 1$, we can always convert them to expressions involving
only logarithms (up to terms of order $1/\Lambda$). This will be
accomplished by repeated use of identities such as \cite{Lewin:81}:
\begin{eqnarray}
\Li_n(1)&=&\zeta(n)\;,\\
\Li_2(-x)+\Li_2\left(-\frac 1 x\right) &=& -\frac{\pi^2}{6}-\frac 1
2 \ln^2(x)\;, \quad x>0\; \label{inv2}\\
\Li_2(x)+\Li_2\left(\frac 1 x\right) &=& \frac{\pi^2}{3} -i\pi \ln x
-\frac 1 2 \ln^2(x)\;, \quad x>1\;\\
\Li_3(-x)-\Li_3\left(-\frac 1 x\right) &=& -\frac{\pi^2}{6}\ln x
-\frac 1 6 \ln^3(x)\;, \quad x>0\;\label{inv3}\\
\Li_3(x)-\Li_3\left(\frac 1 x\right) &=& \frac{\pi^2}{3} \ln x
-i\frac{\pi}{2} \ln^2(x) -\frac 1 6 \ln^3(x)\;, \quad x>1\;.
\end{eqnarray}
The most important consequence of the form of these equations is the
appearance of terms linear in $\lambda \sim \ln \Lambda$, because
these lead to contributions to the $\beta$-function.

\begin{figure}[h]
\begin{center}
\psfig{figure=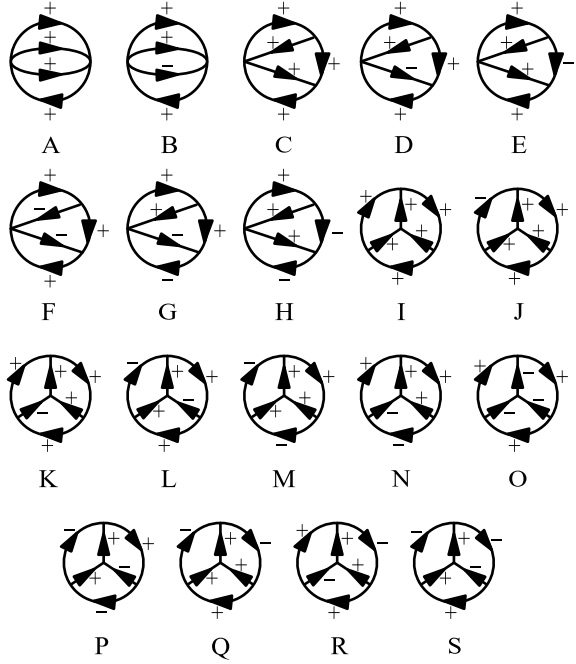} \caption{The basic
non-trivial diagrams at three loops. \label{fig:3loop_basic}}
\end{center}
\end{figure}

We saw in the previous section that, in the end we only needed to
perform one non-trivial integral to be able to calculate the entire
two loop contribution. At three loops, we would of course also like
to narrow down the number of necessary integrals to perform
explicitly as much as possible. The divergences of diagrams A - D of
figure (\ref{fig:3loop}) can be computed from those at two loops
(and are, as already stated, irrelevant for the $\beta$-function).
To be able to compute the contributions from the other diagrams, E -
H of figure (\ref{fig:3loop}), it turns out that we still have to
perform 19 integrals in total. For completeness, we list the
divergent part of these integrals below.
\begin{eqnarray}
{\cal I}_A &=& \int d\tau \; \Delta_+(\tau)\Delta_+(\tau)
\Delta_+(\tau) \partial_{\tau}\Delta_-(\tau) = \frac{i}{8} \left(
\lambda^3 - \frac{\lambda}{2} \right)\\
{\cal I}_B &=& \int d\tau \; \Delta_+(\tau)\Delta_+(\tau)
\Delta_-(\tau) \partial_{\tau}\Delta_-(\tau) = \frac{i}{12}
\lambda^3\\
{\cal I}_C &=& \int d\tau_1 d\tau_2 \; \Delta_+(1)\Delta_+(1+2)
\Delta_+(2) \partial_1\Delta_-(1) \partial_2\Delta_-(2) =
-\frac{1}{8} \left(
\lambda^3 - \frac{\lambda}{3} \right)\\
{\cal I}_D &=& \int d\tau_1 d\tau_2 \; \Delta_+(1)\Delta_+(1+2)
\Delta_-(2) \partial_1\Delta_-(1) \partial_2\Delta_-(2) =
-\frac{1}{16} \left(
\lambda^3 - \frac{\lambda}{2} \right)\\
{\cal I}_E &=& \int d\tau_1 d\tau_2 \; \Delta_+(1)\Delta_-(1+2)
\Delta_+(2) \partial_1\Delta_-(1) \partial_2\Delta_-(2) =
-\frac{1}{24} \left(
\lambda^3 - \frac{\lambda}{2} \right)\\
{\cal I}_F &=& \int d\tau_1 d\tau_2 \; \Delta_-(1)\Delta_+(1+2)
\Delta_-(2) \partial_1\Delta_-(1) \partial_2\Delta_-(2) =
-\frac{1}{24}
\lambda^3 \\
{\cal I}_G &=& \int d\tau_1 d\tau_2 \; \Delta_+(1)\Delta_+(1+2)
\Delta_-(2) \partial_1\Delta_-(1) \partial_2\Delta_+(2) =
\frac{1}{16} \left(
\lambda^3 - \frac{\lambda}{6} \right)\\
{\cal I}_H &=& \int d\tau_1 d\tau_2 \; \Delta_-(1)\Delta_+(1+2)
\Delta_-(2) \partial_1\Delta_-(1) \partial_2\Delta_+(2) =
-\frac{1}{48} \left( \lambda^3 - \frac{\lambda}{2} \right)
\end{eqnarray}
\begin{eqnarray}
{\cal I}_I &=& \int d\tau_1 d\tau_2 d\tau_3 \;
\Delta_+(1+2)\Delta_+(1+2+3) \Delta_+(2+3) \partial_1\Delta_-(1)
\partial_2\Delta_-(2) \partial_3\Delta_-(3)\nonumber\\
&=& -\frac{i}{8} \left(
\lambda^3 - \frac{\lambda}{2} \right)\\
{\cal I}_J &=& \int d\tau_1 d\tau_2 d\tau_3 \;
\Delta_+(1+2)\Delta_+(1+2+3) \Delta_-(2+3) \partial_1\Delta_-(1)
\partial_2\Delta_-(2) \partial_3\Delta_-(3)\nonumber\\
&=& -\frac{i}{16} \left(
\lambda^3 - \frac{\lambda}{2} \right)\\
{\cal I}_K &=& \int d\tau_1 d\tau_2 d\tau_3 \;
\Delta_+(1+2)\Delta_-(1+2+3) \Delta_+(2+3) \partial_1\Delta_-(1)
\partial_2\Delta_-(2) \partial_3\Delta_-(3)\nonumber\\
&=& -\frac{i}{24} \left(
\lambda^3 - \lambda \right)\\
{\cal I}_L &=& \int d\tau_1 d\tau_2 d\tau_3 \;
\Delta_-(1+2)\Delta_+(1+2+3) \Delta_-(2+3) \partial_1\Delta_-(1)
\partial_2\Delta_-(2) \partial_3\Delta_-(3)\nonumber\\
&=& -\frac{i}{24} \left(
\lambda^3 - \frac{\lambda}{2} \right)\\
{\cal I}_M &=& \int d\tau_1 d\tau_2 d\tau_3 \;
\Delta_+(1+2)\Delta_+(1+2+3) \Delta_-(2+3) \partial_1\Delta_-(1)
\partial_2\Delta_-(2) \partial_3\Delta_+(3)\nonumber\\
&=& \frac{i}{48} \left(
\lambda^3 + \frac{\lambda}{2} \right)
\end{eqnarray}
\begin{eqnarray}
{\cal I}_N &=& \int d\tau_1 d\tau_2 d\tau_3 \;
\Delta_+(1+2)\Delta_-(1+2+3) \Delta_+(2+3) \partial_1\Delta_-(1)
\partial_2\Delta_-(2) \partial_3\Delta_+(3)\nonumber\\
&=& \frac{i}{48} \left(
\lambda^3 - \frac{\lambda}{2} \right)\\
{\cal I}_O &=& \int d\tau_1 d\tau_2 d\tau_3 \;
\Delta_+(1+2)\Delta_-(1+2+3) \Delta_-(2+3) \partial_1\Delta_-(1)
\partial_2\Delta_-(2) \partial_3\Delta_+(3)\nonumber\\
&=& \frac{i}{24} \left(
\lambda^3 - \frac{\lambda}{2} \right)\\
{\cal I}_P &=& \int d\tau_1 d\tau_2 d\tau_3 \;
\Delta_-(1+2)\Delta_+(1+2+3) \Delta_-(2+3) \partial_1\Delta_-(1)
\partial_2\Delta_-(2) \partial_3\Delta_+(3)\nonumber\\
&=& 0\\
{\cal I}_Q &=& \int d\tau_1 d\tau_2 d\tau_3 \;
\Delta_+(1+2)\Delta_+(1+2+3) \Delta_-(2+3) \partial_1\Delta_-(1)
\partial_2\Delta_+(2) \partial_3\Delta_-(3)\nonumber\\
&=& 0\\
{\cal I}_R &=& \int d\tau_1 d\tau_2 d\tau_3 \;
\Delta_+(1+2)\Delta_-(1+2+3) \Delta_+(2+3) \partial_1\Delta_-(1)
\partial_2\Delta_+(2) \partial_3\Delta_-(3)\nonumber\\
&=& 0\\
{\cal I}_S &=& \int d\tau_1 d\tau_2 d\tau_3 \;
\Delta_-(1+2)\Delta_+(1+2+3) \Delta_-(2+3) \partial_1\Delta_-(1)
\partial_2\Delta_+(2) \partial_3\Delta_-(3)\nonumber\\
&=& \frac{i}{48} \lambda \,.
\end{eqnarray}
The diagrams corresponding to the integrals are depicted in figure
(\ref{fig:3loop_basic}).

\begin{figure}[h]
\begin{center}
\psfig{figure=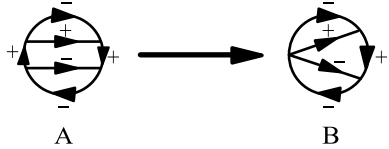} \caption{The integral of
diagram (A) can be related to the one of diagram (B).
\label{fig:ex_diagr}}
\end{center}
\end{figure}

All of these integrals were performed in the same way as the two
loop diagram of appendix (\ref{app 2l}). They all essentially involve
integrals over products of sign functions. Take for instance ${\cal
I}_M$. Using the definition of the $\tau$-space propagators
(\ref{t-space prop}), we find more explicitly that
\begin{eqnarray}
\begin{split}
{\cal I}_M &= \frac{i}{2^6 \pi^3} \int_1^\Lambda \frac{dp}{p}
\int_1^\Lambda \frac{dq}{q} \int_1^\Lambda \frac{dk}{k} \big[
-\epsilon(p+q-\Lambda)
\epsilon(p+q-k-\Lambda) \epsilon(q-k+1) \\
& + \epsilon(p+q-\Lambda) \epsilon(p+q-k-1)
\epsilon(q-k+1)\\
& + \epsilon(p+q-\Lambda) \epsilon(p+q-k-\Lambda) -
\epsilon(p+q-\Lambda) \epsilon(p+q-k-1)\\
& + \epsilon(p+q-k-\Lambda) \epsilon(q-k+1) - \epsilon(p+q-k-1)
\epsilon(q-k+1)\\
& - \epsilon(p+q-k-\Lambda) + \epsilon(p+q-k-1)\big]\,.
\end{split}\label{I_3loop}
\end{eqnarray}
Clearly, this kind of calculation is best done using a computer.
%We computed every term separately using Mathematica and added all terms
%together by hand afterwards to obtain the results listed above. For
%even higher loops it would certainly become necessary to write a
%program that would execute all of these steps at once.
The 19 integrals we have listed above can be seen as the building
blocks out of which every other non-trivial three loop integral can
be obtained quite easily. To illustrate this point, let us look at
the integral corresponding to diagram A of figure
(\ref{fig:ex_diagr}).

The reader may have noticed that this type of diagram did not appear
in the list of `building blocks' of figure (\ref{fig:3loop_basic}).
This is because an integral corresponding to this type of diagram
can always be related to one corresponding to a diagram of the type
of diagram B of figure (\ref{fig:ex_diagr}). This is neatly
illustrated with the case at hand. The integral corresponding to
diagram A of figure (\ref{fig:ex_diagr}) equals
\begin{eqnarray}
\begin{split}
{\cal I} &= \int d\tau_1 d\tau_2 d\tau_3 \; \Delta_-(1)\Delta_-(2)
\Delta_+(1+2+3) \partial_1\Delta_+(1)
\partial_2\Delta_+(2) \partial_3\Delta_-(3)\\
&+ \int d\tau_1 d\tau_2 d\tau_3 \; \Delta_+(1)\Delta_-(2)
\Delta_+(1+2+3) \partial_1\Delta_-(1)
\partial_2\Delta_+(2) \partial_3\Delta_-(3)\,.
\end{split}
\end{eqnarray}
Using the fact that
\begin{eqnarray}
\int d\tau_3 \; \Delta_\pm(1+2+3) \partial_3\Delta_\mp(3) = \pm i
\Delta_\pm(1+2)\,,
\end{eqnarray}
we can perform the integral over $\tau_3$ and arrive at
\begin{eqnarray}
\begin{split}
{\cal I} &= i\int d\tau_1 d\tau_2\; \Delta_-(1)\Delta_-(2)
\Delta_+(1+2) \partial_1\Delta_+(1)
\partial_2\Delta_+(2)\\
&+ i\int d\tau_1 d\tau_2 \; \Delta_+(1)\Delta_-(2) \Delta_+(1+2)
\partial_1\Delta_-(1)
\partial_2\Delta_+(2)\,,
\end{split}
\end{eqnarray}
which indeed corresponds to diagram B of figure
(\ref{fig:ex_diagr}). This diagram does not correspond to any of the
building blocks of figure (\ref{fig:3loop_basic}) either, but
comparison with the ones that do, shows that we can write
\begin{eqnarray}
{\cal I} = i \left( {\cal I}_E + {\cal I}_G \right)\,.
\end{eqnarray}
This illustrates how other integrals one needs to perform can be
related to the building blocks.

The advantage of this method is that it can, in principle, easily be
generalized to higher loops. First of all, superspace techniques can
just as easily be applied to higher loop diagrams. The $\tau$-space
integral one ends up with will always be a generalization of
eqs.~(\ref{I}) and (\ref{I_3loop}). The general procedure explained in
this and the previous appendix to handle these kinds of integrals
can still be applied at higher loops. These integrals will always be
expressible in terms of logarithms and polylogarithms. At $l$ loops,
a general term will be of the form\footnote{Eq. (\ref{generalterm})
only serves as a rough indication of the general form. More
concretely, not all factors of ln need to have the same argument and
there might be more polylogarithms involved in the same term. The
important point is that all powers of ln and all orders of the
polylogarithms involved add up to $l$.}
\begin{eqnarray}
\Li_n \left(r(\Lambda)\right)\;\ln^{l-n}\left(s(\Lambda)\right),
\qquad n \leq l,\label{generalterm}
\end{eqnarray}
where $r$ and $s$ are rational functions of $\Lambda$. The $n$-th
order polylogarithm $\Li_n$ is defined as
\begin{eqnarray}
\Li_n(x) = \int_0^x dz\; \frac{\Li_{n-1}(z)}{z} = \sum_{k=1}^\infty
\frac{x^k}{k^n}\;, \quad \vert x \vert \leq 1\;,\label{polyn}
\end{eqnarray}
and $\Li_2$ is defined in eq.~(\ref{poly2}). The last equality in
eq.~(\ref{polyn}) is only valid for $\vert x\vert \leq 1$, as
indicated, but the integral representation can be used as a
definition of $\Li_n$ on the whole (cut) complex plane. Since we are only
interested in the behavior of eq.~(\ref{generalterm}) for $\Lambda
>> 1$, we can use identities like \cite{Lewin:81}\footnote{For $n=2$
and 3, this expression reduces to eqs.~(\ref{inv2}) and
(\ref{inv3}), respectively.}
\begin{eqnarray}
\Li_n(-x) + (-1)^n \Li_{n}\left(-\frac 1 x \right) &=& -\frac{1}{n!}
\ln^n(x) + 2 \sum_{k=1}^{[n/2]} \frac{\Li_{2k}(-1)}{(n-2k)!}\;
\ln^{n-2k}(x),
\end{eqnarray}
where $[n/2]$ is the greatest integer contained in $n/2$ and
\begin{eqnarray}
\Li_{n}(-1) = \left(2^{1-n}-1\right)\zeta(n),
\end{eqnarray}
to write the asymptotic behavior of (\ref{generalterm}) purely in
terms of $\lambda \sim \ln(\Lambda)$. In the end we still arrive at
a polynomial in $\lambda$, as desired.


\begin{thebibliography}{99}
\bibitem{witten}  E.~Witten,
{\em Bound states of strings and p-branes}, \npb{460}{1996}{35},
\hepth{9510135}.
%%CITATION = HEP-TH 9510135;%%
\bibitem{BI1} E.S. Fradkin and A.A. Tseytlin,
{\em Nonlinear electrodynamics from quantized strings},
\plb{163}{1985}{123}.
%%CITATION = PHLTA,B163,123;%%
\bibitem{direct1} A.A. Tseytlin,
{\em Vector field effective action in the open superstring theory},
\npb{276}{1986}{391} and
\npb{291}{1987}{876}.
%%CITATION = NUPHA,B276,391;%%
\bibitem{BI2} A. Abouelsaood, C. Callan, C. Nappi and S. Yost,
{\em Open strings in background gauge fields},
\npb{280}{1987}{599}.
%%CITATION = NUPHA,B280,599;%%
\bibitem{klaus} K.~Behrndt, {\em Untersuchung der Weyl-Invarianz im verallgemeinterten $ \sigma %%@
$-Modell f\"ur offene Strings}, Dissertation zur Erlangung des akademischen Grades doctor rerum %%@
naturalium, Humboldt-Universit\"at zu Berlin, 1990.
\bibitem{BI3} R.G. Leigh,
{\em Dirac-Born-Infeld action from Dirichlet sigma model},
\mpla{4}{1989}{2767}.
%%CITATION = MPLAE,A4,2767;%%
\bibitem{BI4}
M.~Cederwall, A.~von Gussich, B.~E.~W.~Nilsson and A.~Westerberg,
{\em The Dirichlet super-three-brane in ten-dimensional type-IIB supergravity},
\npb{490}{1997}{163}, \hepth{9610148}.
%%CITATION = HEP-TH 9610148;%%
\bibitem{BI5} M.~Aganagic, C.~Popescu and J.~H.~Schwarz,
{\em D-brane actions with local kappa symmetry},
\plb{393}{1997}{311}, \hepth{9610249}
%%CITATION = HEP-TH 9610249;%%
and {\em Gauge-invariant and gauge-fixed D-brane actions},
\npb{495}{1997}{99}, \hepth{9612080}.
%%CITATION = HEP-TH 9612080;%%
\bibitem{BI6} M.~Cederwall, A.~von Gussich, B.~E.~W.~Nilsson, P.~Sundell and A.~Westerberg,
{\em The Dirichlet super p-branes in ten-dimensional type IIA and IIB supergravity},
\npb{490}{1997}{179}, \hepth{9611159}.
%%CITATION = HEP-TH 9611159;%%
\bibitem{BI7} E.~Bergshoeff and P. K.~Townsend,
{\em Super D-branes},
\npb{490}{1997}{145}, \hepth{9611173}.
%%CITATION = HEP-TH 9611173;%%
\bibitem{andreevtseytlin} O.D.~Andreev and A.A.~Tseytlin, {\em Partition function representation
for the open
superstring effective action: cancellation of M\"obius infinities and derivative corrections to
Born-Infeld Lagrangian}, \npb{311}{1988}{205}.
%%CITATION = NUPHA,B311,205;%%
\bibitem{wyllard} N.~Wyllard, {\em Derivative corrections to D-brane actions with constant
background fields}, \npb{598}{2001}{247}, \hepth{0008125}.
%%CITATION = HEP-TH 0008125;%%
\bibitem{cornalba}
L.~Cornalba, {\em The general structure of the non-abelian Born-Infeld action},
\atmp{4}{2002}{1259}, \hepth{0006018}.
%%CITATION = HEP-TH 0006018;%%
%\cite{Collinucci:2002gd}
\bibitem{Collinucci:2002gd}
A.~Collinucci, M.~de Roo and M.~G.~C.~Eenink,
{\em Derivative corrections in 10-dimensional super-Maxwell theory},
\jhep{0301}{2003}{039}, \hepth{0212012}.
  %%CITATION = HEP-TH 0212012;%%
\bibitem{goteborg} E.~Bergshoeff, M.~Rakowski and E.~Sezgin,
{\em Higher-derivative super Yang-Mills theories},
\plb{185}{1987}{371};
%%CITATION = PHLTA,B185,371;%%
M. Cederwall, B.E.W. Nilsson and D. Tsimpis,
{\em The structure of maximally supersymmetric Yang-Mills
theory: constraining higher-order corrections},
\jhep{0106}{2001}{034}, \hepth{0102009}
%%CITATION = HEP-TH 0102009;%%
and {\em D=10 super Yang-Mills at $\ap{2}$}, \jhep{0107}{2001}{042},
\hepth{0104236}.
%%CITATION = HEP-TH 0104236;%%
%\cite{Tseytlin:1997cs}
\bibitem{Tseytlin:1997cs}
  A.~A.~Tseytlin,
{\em On non-abelian generalisation of the Born-Infeld action in string
theory}, \npb{501}{1997}{41}, \hepth{9701125}.
  [arXiv:hep-th/9701125].
  %%CITATION = HEP-TH 9701125;%%
%\cite{Hashimoto:1997gm}
\bibitem{Hashimoto:1997gm}
  A.~Hashimoto and W.~I.~Taylor,
{\em Fluctuation spectra of tilted and intersecting D-branes from the
Born-Infeld action}, \npb{503}{1997}{193}, \hepth{9703217}.
  %%CITATION = HEP-TH 9703217;%%
%\cite{Denef:2000rj}
\bibitem{Denef:2000rj}
  F.~Denef, A.~Sevrin and J.~Troost,
{\em Non-Abelian Born-Infeld versus string theory}, \npb{581}{2000}{135},
\hepth{0002180}.
  %%CITATION = HEP-TH 0002180;%%
\bibitem{bdrs} E.A.~Bergshoeff, M.~de Roo and A.~Sevrin, unpublished.
\bibitem{direct} D. J. Gross and E. Witten,
{\em Superstring modifications of Einstein's equations},
\npb{277}{1986}{1}.
%%CITATION = NUPHA,B277,1;%%
\bibitem{alpha3} P. Koerber and A. Sevrin, {\em The non-abelian open superstring
effective action through order $\alpha'{}^3$}, \jhep{0110}{2001}{003}, \hepth{0108169}.
%%CITATION = HEP-TH 0108169;%%
%\cite{Barreiro:2005hv}
\bibitem{Barreiro:2005hv}
L.~A.~Barreiro and R.~Medina,
{\em 5-field terms in the open superstring effective action},
\jhep{0503}{2005}{055}, \hepth{0503182}.
  %%CITATION = HEP-TH 0503182;%%
\bibitem{stieberger}
 D.~Oprisa and S.~Stieberger,
{\em Six gluon open superstring disk amplitude, multiple hypergeometric series
and Euler-Zagier sums}, \hepth{0509042}.
%%CITATION = HEP-TH 0509042;%%
\bibitem{lies} L.~De Foss\'e, P.~Koerber and A.~Sevrin,
{\em The uniqueness of the Abelian Born-Infeld action},
\npb{603}{2001}{413}, \hepth{0103015}.
%%CITATION = HEP-TH 0103015;%%
\bibitem{alpha4} P. Koerber and A. Sevrin, {\em The non-abelian D-brane effective action through %%@
order
$ \alpha '{}^4$}, \jhep{0210}{2002}{046}, \hepth{0208044}.
%%CITATION = HEP-TH 0208044;%%
\bibitem{testalpha4}
A.~Sevrin and A.~Wijns,
{\em Higher order terms in the non-abelian D-brane effective action and  magnetic background %%@
fields},
\jhep{0308}{2003}{059}, \hepth{0306260}.
%%CITATION = HEP-TH 0306260;%%
%\cite{Grasso:2002wb}
\bibitem{Grasso:2002wb}
  D.~T.~Grasso,
{\em Higher order contributions to the effective action of N = 4 super
Yang-Mills}, \jhep{0211}{2002}[012], \hepth{0210146}.
  %%CITATION = HEP-TH 0210146;%%
%\cite{Refolli:2001df}
\bibitem{Refolli:2001df}
  A.~Refolli, A.~Santambrogio, N.~Terzi and D.~Zanon,
{\em $F^5$ contributions to the nonabelian Born Infeld action from a
supersymmetric Yang-Mills five-point function}, \npb{613}{2001}{64}, erratum
\npb{648}{2003}{453}, \hepth{0105277}.
  %%CITATION = HEP-TH 0105277;%%
\bibitem{Grisaru1}
 M.~T.~Grisaru, A.~E.~M.~van de Ven and D.~Zanon,
{\em Four Loop Divergences For The N=1 Supersymmetric Nonlinear Sigma Model In
Two-Dimensions}, \npb{277}{1986}{409}.
%%CITATION = NUPHA,B277,409;%%
%\cite{Grisaru:1986dk}
\bibitem{Grisaru2}
M.~T.~Grisaru, A.~E.~M.~van de Ven and D.~Zanon,
{\em Two-Dimensional Supersymmetric Sigma Models On Ricci Flat Kahler Manifolds
Are Not Finite}, \npb{277}{1986}{388}.
%%CITATION = NUPHA,B277,388;%%
%\cite{Bordalo:2004xg}
\bibitem{Bordalo:2004xg}
  P.~Bordalo, L.~Cornalba and R.~Schiappa,
  {\em Towards quantum dielectric branes: Curvature corrections in abelian  beta
  function and nonabelian Born-Infeld action}, \npb{710}{2005}{189}, \hepth{0409017}
    %%CITATION = HEP-TH 0409017;%%
%\cite{Koerber:2003ef}
\bibitem{Koerber:2003ef}
P.~Koerber, S.~Nevens and A.~Sevrin,
{\em Supersymmetric non-linear sigma-models with boundaries revisited}, \jhep{11}{2003}{066},
\hepth{0309229}.
%%CITATION = HEP-TH 0309229;%%
%\cite{Corrigan:1982th}
\bibitem{Corrigan:1982th}
  E.~Corrigan, C.~Devchand, D.~B.~Fairlie and J.~Nuyts,
{\em First Order Equations For Gauge Fields In Spaces Of Dimension Greater Than
 Four}, \npb{214}{1983}{452}.
    %%CITATION = NUPHA,B214,452;%%
\bibitem{duy} K.~Uhlenbeck and S.-T.~Yau,
{\em On the existence of hermitian Yang-Mills connections on stable vectorbundles},
{\em Comm. Pure Appl. Math.} {\bf 39}
(1986) 257 and {\em A note on our previous paper: on the existence of hermitian Yang-Mills
connections on
stable vectorbundles}, {\em Comm. Pure Appl. Math.} {\bf 42} (1989) 703; S.K. Donaldson,
{\em Infinite determinants, stable bundles and curvature},
{\em Duke Math. J.} {\bf 54} (1987) 231; see also chapter 15 in the second volume M.B. Green,
J.H. Schwarz and
E. Witten, {\em Superstring theory}, Cambridge University Press 1986.
\bibitem{koerber:thesis} P.~Koerber, {\em Abelian and Non-abelian D-brane Effective
Actions}, \forp{52}{2004}{871}, \hepth{0405227};
%%CITATION = HEP-TH 0405227;%%
\bibitem{wip} A. Sevrin, W. Troost and A. Wijns, work in progress.
\bibitem{dorn}
H.~Dorn and H.~J.~Otto, {\em On T-duality for open strings in general abelian and nonabelian %%@
gauge field
backgrounds}, \plb{381}{1996}{81}, \hepth{9603186};
%%CITATION = HEP-TH 9603186;%%
H.~Dorn, {\em Nonabelian gauge field dynamics on matrix D-branes}, \npb{494}{1997}{105}, %%@
\hepth{9612120}.
%%CITATION = HEP-TH 9612120;%%
\bibitem{Gates:1983nr}
S.~J.~Gates, M.~T.~Grisaru, M.~Rocek and W.~Siegel, {\em Superspace,
Or One Thousand And One Lessons In Supersymmetry}, Front.\ Phys.\
{\bf 58} (1983) 1, \hepth{0108200};
%%CITATION = HEP-TH 0108200;%%
\bibitem{Lewin:81}
L.~Lewin, {\em Polylogarithms and Associated Functions}, Elsevier
North Holland, Inc., 1981.
\end{thebibliography}
\end{document}